\documentclass[prl,twocolumn]{revtex4}
\usepackage{graphicx}
\begin{document}
\title{Electronic-vibrational coupling in single-molecule devices}
\author{Vivek Aji}
\affiliation{Department of Physics, University of California, Berkeley, CA 94720}
\affiliation{Materials Sciences Division, Lawrence Berkeley National Laboratory, Berkeley, CA 94720}
\author{Joel E. Moore}
\affiliation{Department of Physics, University of California, Berkeley, CA 94720}
\affiliation{Materials Sciences Division, Lawrence Berkeley National Laboratory, Berkeley, CA 94720}
\author{C. M. Varma}
\affiliation{Bell Labs Lucent Technologies, Murray Hill, NJ 07974}
\date{\today}

\begin{abstract}
Experiments studying vibrational effects on electronic transport through single molecules have observed several seemingly inconsistent behaviors, ranging from up to 30 harmonics of a vibrational frequency in one experiment, to an absence of higher-harmonic peaks in another.  We study the different manifestations of electronic-vibrational coupling in inelastic and elastic electron transport through single molecules.  For the case of inelastic transport, higher harmonics are shown to be damped by additional small factors beyond powers of the electron-vibration coupling constant $\lambda$.   Two mechanisms greatly increase the size of secondary peaks in inelastic transport:
coupling between electron transport and spatial motion of the molecule, and the ``pumping'' of higher vibrational modes of the molecule when vibrational excitations do not completely relax between electron transits.
\end{abstract}
\maketitle

Electronic devices made from single molecules offer both a new means of observing fundamental physics, such as the Kondo effect~\cite{park,mceuen}, and the potential
to circumvent the device size limitations of silicon electronics~\cite{reed}.  Transport in molecular electronic devices is strongly affected by several mechanisms that are negligible in ordinary semiconductor electronics.  An example is the observation of molecular vibrational modes in DC transport measurements on single-molecule devices made from fullerenes~\cite{alivisatos} and other organic molecules~\cite{zhitenev}.  Coupling between molecular vibrations and the electronic states responsible for transport was also proposed as an explanation for the strong temperature dependence of the current through single molecules~\cite{williams}.

There are also experiments, however, in which the electronic-vibrational coupling appears quite weak, as in the STM measurements on an adsorbed molecule reported in~\cite{ho}.  This paper discusses the most important mechanisms for electronic-vibrational effects in transport through single molecules.  The fact that different experiments probe different mechanisms explains much of the observed variation.  The apparent coupling measured in an experiment depends on whether elastic or inelastic tunneling is being probed; whether transport electrons ``pump'' the molecule into higher vibrational states; and whether the molecule's motion affects its coupling to the leads.  Our calculations are based on single-level Einstein-type models for electron-phonon coupling, and in what follows both ``phonon'' and ``vibrational mode'' describe the vibrational states of the active molecule.

The chief simplifying assumptions in our calculations are that the system is at zero temperature ($T=0$) and off resonance, so that the conductance per molecule $g \ll \frac{e^2}{h}$.  For the systems under consideration the phonon energy $\hbar \omega$ and bare level width $\Gamma$ are both typically larger than the experimental temperature.  Under the assumption that incident electrons reach the molecule in its ground state, the vibrational modes of the molecule
appear first as inelastic {\it steps} in the differential conductance (when an electron passing through leaves behind one or more phonons), and then as subpeaks of resonant-tunneling (elastic) {\it peaks}.  The height of inelastic steps is shown to fall off much more rapidly with phonon number $n$ than the weight of elastic subpeaks, explaining the lack of higher-harmonic peaks in~\cite{ho}.  This results from an orthogonality constraint and hence should hold beyond the simple Einstein model used here.

The transport of electrons becomes more complex when the phonons excited by one electron's passing through do not have time to relax before the next electron arrives: there is a nonthermal steady-state of phonon occupancies on the molecule, and high phonon-number states are ``pumped'' as soon as the bias voltage is larger than the phonon energy.  We also consider a model where electron hopping on/off the molecule is modified by vibrational excitations, motivated by the experiments~\cite{alivisatos,zhitenev}.  The final part of the paper discusses which predictions of the idealized Einstein-type models can be expected to apply more generally.  Quantities such as overlap integrals and electron-phonon coupling strength, which here are taken as inputs, can be calculated {\it ab initio} with moderate accuracy~\cite{diventra} for specific molecular systems.

The Einstein Hamiltonian for an electronic level of energy $\epsilon_d$ and a single vibrational mode of energy $\hbar \omega$ is
\begin{equation}
H_d = \hbar \omega b^\dagger b + \epsilon_d c^\dagger c +
\lambda \hbar \omega c^\dagger c (b+b^\dagger).
\end{equation}
This Hamiltonian conserves electron number: the zero-electron state with $p$ phonons, $|0,p\rangle$, has energy $p \hbar \omega$.  The one-electron ground state can
be found by a simple canonical transformation~\cite{mahan} and has energy
${\tilde \epsilon}_d \equiv \epsilon_d-\lambda^2 \hbar \omega$.  It is annihilated by $b+\lambda$ and is a superposition of states of different phonon number:
\begin{equation}
\langle 0,n|c|1,0\rangle = {\lambda^n e^{-\lambda^2/2}\over \sqrt{n!}}.
\end{equation}
Straightforward algebra gives the matrix elements between general
zero- and one-particle eigenstates:
\begin{eqnarray}
\langle 0,n|c|1,k \rangle = \langle n|{1 \over \sqrt{k!}} {(\lambda+b^\dagger)^k e^{-\lambda^2/2}\sum_m {(-\lambda)^m\over
\sqrt{m!}} |m\rangle }\cr
= \sum_{l+m = n}\lambda^{k-l} \left({k \atop l}\right) \sqrt{n!\over m!}{(-\lambda)^m e^{-\lambda^2/2}\over\sqrt{m!k!}}\cr
= e^{-\lambda^2/2} \sum_{l=0}^{{\rm min}(k,n)} {\lambda^{k+n-2l}(-1)^{n-l}
\sqrt{k!n!} \over (n-l)!(k-l)!l!}
\end{eqnarray}

To describe an idealized single-molecule device, the hopping of electrons on and off the molecule must be added:
\begin{eqnarray}
H_{\rm tot} &=& H_{L1} + H_{L2} + H_d + \sum_{k1} (t_{k1} c_d^\dagger c_{k1} + {\rm
h.c.}) + \cr
&&\sum_{k2} (t_{k2} c_d^\dagger c_{k2} + {\rm h.c.}).
\end{eqnarray}
Here $H_{L1},H_{L2}$ are the Hamiltonians of the left and right leads, and $t_{k1}$ is the overlap integral between state $k1$ and the molecular level.  This hopping model is appropriate when the molecular level is weakly coupled, but the low conductance of current devices suggests that this is a valid assumption~\cite{schonex}.  It is assumed below that the leads are essentially uniform in energy over the ranges probed in an experiment, and that the lead densities of states $\rho_{L,R}$ and hopping elements $V^{L,R}_d$ are energy-independent.
The left and right level widths are
$\gamma_{L,R} = 2 \pi \rho_{L,R} |V^{L,R}_d|^2$.

Steps in differential conductance appear in the Einstein model when the bias voltage is a multiple of the phonon energy.  When the bias voltage $V \geq n \hbar \omega$, an electron passing through the molecule can excite $n$ phonons.  We find that the step height falls off very rapidly with $n$ for reasonable parameters because of an orthogonality requirement, which may explain the absence of $n>1$ peaks in inelastic STM spectroscopy of single molecules.~\cite{ho} We start by assuming that between electron transits, any phonons which may have been excited by the first electron decay without electron transfer before the second electron arrives.  Later this assumption will be weakened, with dramatic effects on the step strength.  In the limit of rapid phonon decay, each electron reaches the molecule in its ground state $|0,0\rangle$.  The current for
${\tilde \epsilon}_d > 2 V$ occurs via virtual excitation of the molecule:
\begin{eqnarray}
I &=& \sum_{n=0}^{n<e V/ \hbar \omega}I_{0\rightarrow n} \cr
&=& {2 e \Gamma_L \Gamma_R \over 2 \pi h} \sum_{n=0}^{n<e V/ \hbar \omega} \int_{-eV/2 + n \hbar \omega}^{eV/2} A(n,E) \,dE,\cr
A(n,E)&=&\left(\sum^\infty_{k=0} {\langle 0,n|c|1,k\rangle\langle 1,k|c^\dagger|0,0\rangle\over
{\tilde \epsilon}_d + k \hbar \omega-E}\right)^2.
\label{currform}
\end{eqnarray}
\def\ted{{\tilde \epsilon}_d}
When $eV$ crosses $n \hbar \omega$, the contribution of the new channel $I_{0 \rightarrow n}$ is initially zero.  $I$ is continuous and
$dI/dV$ has steps at $eV = n \hbar \omega$ for zero temperature, which broaden with increasing temperature but remain visible for $kT<\hbar \omega$.  The magnitude of these steps can be found explicitly: define the dimensionless step height
\begin{equation}
K_n \equiv {h \over 2 e^2}
\left({dI \over dV}|_{e V=n \hbar \omega^+}-
{dI \over dV}|_{e V=n \hbar \omega^-} \right)
\end{equation}
and the function $f(x) \equiv (-\lambda)^{-x}\left[\Gamma(-x)-
\Gamma(-x,-\lambda)\right] = \sum_{n=0}^\infty {\lambda^n \over n! (x + n)}.$
Then
\begin{eqnarray}
K_1 =& {\Gamma_L \Gamma_R \over 2 \pi (\hbar \omega)^2} \lambda^2
e^{-2 \lambda^2}\Big(f(\ted/\hbar \omega-{\textstyle \frac{1}{2}}) -\cr
&f(\ted/\hbar \omega+{\textstyle \frac{1}{2}})\Big)^2\cr
K_2 =& {\Gamma_L \Gamma_R \over 4 \pi (\hbar \omega)^2} \lambda^4
e^{-2 \lambda^2}
\Big(f(\ted/\hbar \omega-1) -\cr
&2 f(\ted/\hbar \omega)+f(\ted/\hbar \omega+1)\Big)^2
\end{eqnarray}
and the pattern of discrete derivatives continues to higher steps.  The first two step heights are plotted in Fig. 1, normalized by the smooth part of the
differential conductance at $eV = \hbar \omega$,
\begin{eqnarray}
G =& {dI \over dV}|_{e V=\hbar \omega^-}={2 e^2 \over h} {\Gamma_L \Gamma_R \over 2 \pi (\hbar \omega)^2} {e^{-2 \lambda^2} \over 2} \times \cr
&\left(f(\ted/\hbar \omega -1/2)^2 + f(\ted/\hbar \omega + 1/2)^2\right)
\end{eqnarray}

\begin{figure}
\includegraphics[width=3.0in]{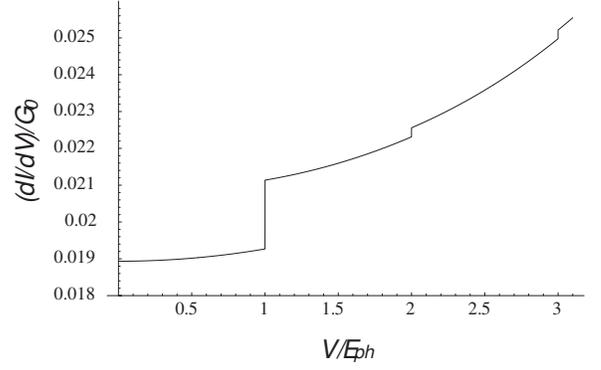}
\caption{Inelastic steps in differential conductance: $(dI / dV)$ in units of $G_0 = 2e^2 \Gamma_L \Gamma_R/(2 \pi h (\hbar \omega)^2)$, versus $V(\hbar \omega)^{-1}$, for $\lambda=2$, ${\tilde \epsilon}_d = 3.75 \hbar \omega$.  Even at this strong coupling, higher steps are much smaller than the first step because of additional factors of $(\hbar \omega / {\tilde \epsilon}_d)$.}
\label{figone}
\end{figure}

The main qualitative prediction of the above is a simple consequence of the orthogonality of phonon states: inelastic steps are weak (compared to inelastic peaks) because the sum in (\ref{currform}) vanishes when $(\hbar \omega /{\tilde \epsilon}_d) \rightarrow 0$.
For small $\lambda$ and small $\hbar \omega / \ted$, the height of the $n$th step decreases with $n$ as $(\lambda \hbar \omega/\ted)^{2n}$.  Even for $\lambda=2$ the
$n>1$ steps are very small, as seen in Fig.~\ref{figone}.  The elastic tunneling peaks in the Einstein model (discussed below) fall off only as $\lambda^n$; the additional factor $\hbar \omega / \ted$ is absent.

The subsidiary steps can be
much larger if the phonon decay time is slow compared to the current, as then phonons
are predominantly excited and relaxed by the transport electrons.
Before considering this ``pumping'' mechanism, we briefly review the effect of phonons on {\it elastic} peaks.  Under the weak-coupling assumption of this paper, the differential conductance is simply proportional to the electron density of states on the molecule.  Start by fixing $\lambda = 0$ so that there is one resonant tunneling peak, at $eV/2 = \ted$.  The electron density of states is Lorentzian, and $\Delta \rho$ per spin integrates to one electron:
\begin{equation}
\Delta \rho(E) = {\Gamma_L + \Gamma_R \over \pi (E^2 + (\Gamma_L + \Gamma_R)^2/4)}.
\end{equation}
The differential conductance is then~\cite{meir}
\begin{equation}
{dI \over dV} = {2 e^2 \over \hbar} {\Gamma_L \Gamma_R \over (\Gamma_L + \Gamma_R)} {\Delta \rho(eV/2) \over 2},
\end{equation}
and the total peak weight is $W_0 = 2 e \Gamma_L \Gamma_R / \hbar (\Gamma_L + \Gamma_R)$.

The above results generalize quite simply to the Einstein model as long as the level broadening $\Gamma_L, \Gamma_R \ll \hbar \omega$.  The $n$th resonant peak is a
Lorentzian centered on $eV/2 = \ted+n \hbar \omega$ and has
width $(\Gamma_L + \Gamma_R)/2$, as before, and weight $W=W_0 \sum_n |\langle 0,0|c|1,n\rangle|^2$
under the standard resonant tunneling assumptions for an interacting level\cite{glazman,ng,meir}.  Hence the total weight in all the differential conductance peaks is the same as before; the broadened lineshape generated by electron-phonon coupling is discussed in~\cite{glazman2}.  These elastic subpeaks of a resonant tunneling peak have been observed in~\cite{alivisatos}.  For small $\lambda$, the peaks fall off as $\lambda^{2n}$; note
that the factor $(\hbar \omega / \ted)^{2n}$ which was present for the step heights in the inelastic case is now absent.

The current through a single-molecule device is strongly modified when vibrational excitations on the molecule do not relax completely between electron transits.  There is a crossover between two regimes with increasing current: low current is described by 
the preceding results, while at high current the vibrational excitations are predominantly excited and relaxed by transport electrons, rather than by decaying directly into the leads.  Calculating the crossover requires knowing the rate of vibrational excitations from decay into the bulk without electron transport, which depends sensitively on the details of the molecule-lead contacts.  The phonon lifetime for nanoparticles on surfaces is discussed in~\cite{patton}.

We consider for brevity the limit where vibrational excitations on the molecule are created and annihilated only by transport electrons.  This situation is expected to be more easily realized using molecules on surfaces~\cite{ho}; molecules strongly contacted to leads will have rapid decay of phonons into the bulk leads.  The rate equations obtained in this limit can be easily modified to include vibrational relaxation without electrons; there is a smooth crossover from decay entirely through transport electrons to decay entirely through phonon relaxation into the bulk leads.  The matrix element for a typical process in second-order perturbation theory is
\begin{equation}
r^{L\rightarrow R}_{j\rightarrow k} (E_L) = t_L t_R \sum_n {\langle 0,j| c|1,n \rangle \langle 1,n|c^\dagger|0,k>
\over \epsilon_d - \lambda^2 \hbar \omega- E_L + (n-j)\hbar \omega}.
\end{equation}
This notation means that the electron moves from $L$ to $R$ and starts with energy
$E_L$, in the process changing the vibrational state from $j$ to $k$.

The $j \neq k$ terms vanish if $\omega/(\ted-E_L) \rightarrow 0$,
since then the denominator is constant and the sum evaluates to
$\delta_{jk}/(\ted - E_L)$.  The pumping mechanism depends on having a vibrational mode frequency not too much smaller than the energy scale of the electronically excited state.  On the other hand, a large vibrational frequency means that peaks only appear at higher voltages, so there is a window of vibrational
frequencies which will have the most significant effects.  This may partly explain the experimental observation~\cite{zhitenev} that only one or a few vibrational modes seem to show up strongly in transport.   As required by unitarity, $r^{L\rightarrow R}_{j\rightarrow k} (E_L)=
r^{R\rightarrow L}_{j\rightarrow k} (E_L+(k-j) \hbar \omega)$.

The total rate for the molecule to change from state $j$ to state $k$, $k>j$, is then
\begin{eqnarray}
\Gamma^{L\rightarrow R}_{j \rightarrow k} &=& \int^{\mu_L}_{\mu_R+\hbar \omega (k-j)}  \,dE_L\,\Gamma_L \Gamma_R\cr
&&\left(\sum_n {\langle 0,j| c|1,n \rangle \langle 1,n|c^\dagger|0,k>
\over \epsilon_d - \lambda^2 \hbar \omega- E_L + (n-j)\hbar \omega}
\right).
\end{eqnarray}
Here $\Gamma_{L,R} = 2 \pi |t_{L,R}|^2 \rho_{L,R}$ and $\rho_{L,R}$ is the density of states (assumed constant as a function of energy).  In what follows symmetric coupling is assumed: $\Gamma=\Gamma_L = \Gamma_R$.  Similar equations with more terms apply for the relaxation rates, since there are more ways the molecule can relax.  These rate equations are similar to those for spin excitations in~\cite{wegewijs} but with many more states involved.

The rate equations for virtual (second-order) tunneling cease to apply once sequential (first-order) tunneling becomes energetically allowed.  Sequential tunneling effectively
closes the rate equations, since high vibrational levels of the molecule
are quickly relaxed once sequential tunneling becomes possible
($n \hbar \omega + \mu_L \geq {\tilde \epsilon}_d$).  This is again a consequence of our assumption that the molecule is weakly contacted, i.e., $\Gamma \gg \Gamma^2 / \epsilon_d$.  As an example of the pumping effect, we calculate that for $\lambda = 1, {\tilde \epsilon}_d = 5.5$, the first step is enhanced by a factor $2.49$ relative to no pumping, and the second step by a factor $7.23$.  These numbers include no transport by sequential tunneling from highly excited states; this would further increase the step size in the pumped case.  Now we return to rapid relaxation (no pumping) but consider how spatial motion of the molecule modifies tunneling.

It has been suggested~\cite{alivisatos,zhitenev} that the vibrational modes observed in experiments are related to the center-of-mass oscillation of the molecule. Such molecular motion affects the overlap of the electron wavefunction
on the molecule and the leads, and hence the sequential tunneling rates $\Gamma_L$ and $\Gamma_R$. To study its effect on the peaks in differential conductance,
we use a simple generalization of the Einstein model, where $\Gamma_{L}$ and $\Gamma_{R}$ are now dependent on the
oscillator state $n$. Assume that the
the electronic state is a Gaussian centred on the molecule, and that the electronic states on the leads are
$\psi_{L} = \sqrt{2/r_{0}}e^{-x/r_{0}}$,
$\psi_{R} = \sqrt{2/r_{0}}e^{(d-x)/r_{0}}$
where the leads are at $x = 0$ and $d$, and $r_{0}$ is the spread of the wavefunction. 

\begin{figure}\label{spatial}
\vspace{0.2in}
\includegraphics[width=3.0in]{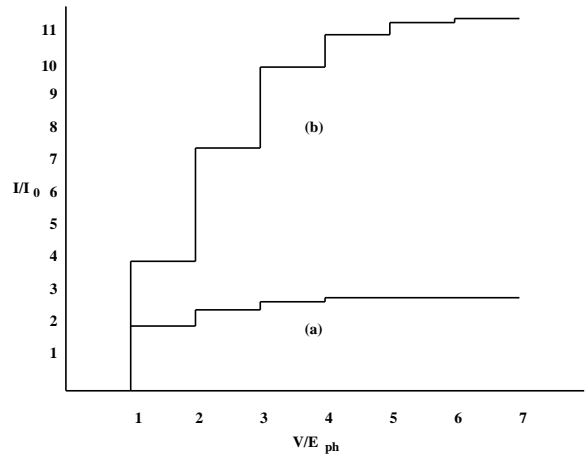}
\caption{Steps in current-voltage due to elastic tunneling process (a) without and (b) with dependence of tunneling rates $\Gamma_L, \Gamma_R$
on vibrational state of the molecule.   In (b), the parameters are $\lambda = 1$ and $l/r_{0} = 1$; in (a), there is no spatial dependence ($l \rightarrow 0$) and again
$\lambda=1$. }
\end{figure}

The wavefunction of the
molecule in the $n$th harmonic oscillator state is
\begin{equation}
\psi_{M}^{n}=(\sqrt{\pi} l 2^{n} n!)^{-1/2} e^{-(x-x_{0})^{2}/2l^{2}}H_{n}((x-x_{0})/l)
\end{equation}
where $l=\sqrt{\hbar/m\omega}$, $H_{n}$ are Hermite polynomials and $x_{0}$ is the mean position of
the molecule.   In a Born-Oppenheimer-type approximation, discussed below, the overlap integrals are obtained and averaged over the probability distribution of the molecule in the $n$th harmonic oscillator state. The resulting tunneling rates are
\begin{eqnarray}\label{eqn_sp_1}
\Gamma_{L} &=& ({4\sqrt{\pi}r_{e} \over {r_{0}}})e^{r_{e}^{2}/r_{0}^{2}}e^{l^{2}/r_{0}^{2}}e^{-2x_{0}/r_{0}}L_{n}({-2l^{2} \over {r_{0}^{2}}})\\ \nonumber
\Gamma_{R} &=& ({4\sqrt{\pi}r_{e} \over {r_{0}}})e^{r_{e}^{2}/r_{0}^{2}}e^{l^{2}/r_{0}^{2}}e^{2x_{0}/r_{0}}e^{2d/r_{0}}L_{n}({-2l^{2} \over {r_{0}^{2}}})
\end{eqnarray}
where $L_{n}$ are Laguerre polynomials and $r_{e}$ is the spread of the electronic state. Note that
for a molecule in a very tight potential, i.e. $l \rightarrow 0$, the results are independent of $n$.
The initial increase in $\Gamma_{L/R}$ with
$n$ results in an increase in the peaks observed in the differential conductance, and depends only on the relative
spread of the wave functions at the leads and that of the harmonic oscillator states.  Fig.~\ref{spatial} shows the resulting current-voltage curves with and without spatial dependence for moderate electron-phonon coupling $\lambda=1$.  The higher harmonic peaks have enhanced weight and show more than seven appreciable steps; these results agree well with the increase in the strength of higher peaks observed in \cite{zhitenev}.

The Born-Oppenheimer approximation that gives (\ref{eqn_sp_1}) is justified in the high-current limit, where a large number of electrons tunnel through the molecule in a single period of the oscillator.  In the case of very low current and temperature (for example, less than one electron in every 100 periods tunnels through the molecule in~\cite{zhitenev}), the analysis is more complicated because it is no longer justified to average over the molecule's probability density.  The finite temperature and molecule-environment interaction in real experiments will help validate Born-Oppenheimer approximation of incoherent molecular motion.

The above results explain, first, why inelastic tunneling experiments often fail to see higher phonon peaks, and second, why elastic tunneling experiments see a large number of peaks even for moderate electron-phonon coupling.  This paper constructed an explicit model of direct coupling between molecular motion and electronic tunneling, which quantitatively explains the current-voltage characteristics seen in experiments~\cite{alivisatos,zhitenev} with reasonable values of the electron-phonon coupling and vibrational amplitude.  Strong enhancement of vibrational effects also occurs when the current is sufficiently high for ``pumping'' of phonons by transport electrons.

The authors wish to thank N. Zhitenev for useful comments.
V. A. and J. E. M. were supported by LBNL grant LDRD-366434 and NERSC.
\bibliographystyle{unsrt}
\bibliography{newphonon}
\end{document}